\documentclass[letterpaper]{article}
 \pdfoutput=1
\usepackage{hyperref}
\usepackage{graphicx}
\usepackage{geometry}
\begin{document}

\title{Determination of the critical points for systems of directed percolation class using machine learning} 
\author{M. Ali Saif$^1$\footnote{masali73@gmail.com}, Bassam M. Mughalles$^2$\footnote{bassammughales@gmail.com}\\
$^1$ Department of Physics,
University of Amran,
Amran,Yemen.\\
$^2$ Department of Physics, University of Sana'a, Sana'a, Yemen.}
\maketitle
\begin{abstract}
Recently, machine learning algorithms have been used remarkably to study the equilibrium phase transitions, however there are only a few works have been done using this technique in the nonequilibrium phase transitions. In this work, we use the supervised learning with the convolutional neural network (CNN) algorithm and unsupervised learning with the density-based spatial clustering of applications with noise (DBSCAN) algorithm to study the nonequilibrium phase transition in two models. We use CNN and DBSCAN in order to determine the critical points for directed bond percolation (bond DP) model and Domany-Kinzel cellular automaton (DK) model. Both models have been proven to have a nonequilibrium phase transition belongs to the directed percolation (DP) universality class. In the case of supervised learning we train CNN using the images which are generated from Monte Carlo simulations of directed bond percolation. We use that trained CNN in studding the phase transition for the two models. In the case of unsupervised learning, we train DBSCAN using the raw data of Monte Carlo simulations. In this case, we retrain DBSCAN at each time we change the model or lattice size. Our results from both algorithms show that, even for a very small values of lattice size, machine can predict the critical points accurately for both models. Finally, we mention to that, the value of the critical point we find here for bond DP model using CNN or DBSCAN is exactly the same value that has been found using transfer learning with a domain adversarial neural network (DANN) algorithm \cite{she}.

\end{abstract}
      

\section{Introduction}
In the field of artificial intelligence, machine learning is a powerful tool which has been attracted a lot of attention today \cite{mi,goo,eng,gor,tag,yam}. Machine learning has been demonstrated its ability and capability in various fields of science and technology, ranging from images classification and speech recognition to natural language processing and autonomous vehicles \cite{car,zde}. Recently, due to its competency to capture features and classifications, machine learning has also been widely employed to study the equilibrium phase transition. Systems such as Ising model \cite{wan,hua}, XY model \cite{zha}, Potts model \cite{li,tan,xc}, Hubbard model \cite{chn}, condensed matters \cite{car1,sc,ven} and quantum phase transition \cite{don,hue,tom} have been studied using machine learning. More recently, machine learning has also been used to study the nonequilibrium phase transition. Wanzhou Zhang et. al. have been applied machine learning with supervised and unsupervised to study the phase transitions of site and bond percolations \cite{zha}. Jianmin Shen et. al. have been used supervised and unsupervised to study the phase transitions of directed percolation in both $(1+1)$ and $(2+1)$ dimensions \cite{she1}. Phase transitions of site percolation and directed percolation have been also studied using transfer learning \cite{she}. In this work we will use supervised and unsupervised learning to study the phase transition for two models which are directed bond percolation model and Domany-Kinzel cellular automaton model. 
  
Machine learning is a field of artificial intelligence that aims to develop computer algorithms and models which can automatically learn patterns from data, improve their performance on the task at hand, and make predictions or decisions without being explicitly programmed to do so \cite{goo,eng}. Machine learning is an attempting to make the computer simulates the function of human brain. In machine learning, we train the computer on a large data set of examples which are known outcomes, and computer uses sophisticated algorithms to learn from this data and generalize its predictions to new, unseen examples. This process involves iteratively adjusting the model's parameters based on feedback from the training data, until the model is able to accurately predict outcomes on test data. 

Supervised learning \cite{gor,tag,she1,van} and unsupervised learning \cite{she1,hu,we,wa} are the two main categories of the machine learning algorithms. The difference between the two algorithms is the kind of training data they work with. Supervised learning involves training a model on a labeled dataset, where each example is associated with a known output or target variable. The goal of the model is to learn a mapping between the input features and the output variable, so that it can make accurate predictions about the output variable for new inputs. Examples of using supervised learning include image classification, language translation, and weather prediction. Unsupervised learning, on the other hand, involves training a model on an unlabeled dataset, where there are no known output variables. Instead, the model is tasked with finding patterns and relationships within the data and clustering similar data points together. It is like letting the model explore and learn autonomously from the data. Examples of using unsupervised learning include discovering topics in text data, identifying groups of customers with similar behavior, and detecting anomalies in network traffic. The third category of machine learning algorithms is called transfer learning which comes from mixing both supervised learning and unsupervised learning.

In the case of phase transition, supervised learning techniques are useful in classifying the phases of systems those undergo a phase transition and determine the critical points, while unsupervised learning algorithms are more suitable for clustering and dimensionality reduction \cite{wan,wet}. There are many of supervised learning algorithms for example linear regression, logistic regression, decision trees, random forest, support vector machines and neural networks. Unsupervised learning algorithms include, hierarchical clustering, K-means clustering, density-based spatial clustering of applications with noise (DBSCAN), expectation-maximization (EM), principal component analysis (PCA), association rule mining and self-organizing maps (SOM). These are just a few examples of the many supervised and unsupervised learning algorithms which are available. The choice of algorithm depends on the specific problem we are trying to solve, the type of data we have and other factors such as computational resources and the need for interpretability. In this work we will use supervised learning techniques with neural networks algorithm and unsupervised learning with density-based spatial clustering of applications with noise algorithm to study the phase transition for systems which undergo a nonequilibrium phase transition belongs to directed percolation class. We will examine the ability of machine learning in identifying the phases of those systems and determine the critical points.

\section{Algorithms Description}
We introduce here a brief description for the supervised learning algorithm and unsupervised learning algorithm which we use them in this work.
\subsection{Supervised Learning Algorithm} 
A convolutional neural network (CNN) is a type of neural network which is inspired by the brain neurons \cite{yam,lec,kr,goh}. It is commonly used for processing data that has a grid pattern, such as image classification, segmentation, object detection and other computer vision tasks. It is particularly effective in handling large spatial data, such as images and video, and are widely used in industry and academia for a variety of applications. Unlike a fully connected neural network where all input features are connected to all hidden layers, CNN is designed to process input images by leveraging the spatial relationships between pixels. CNN typically is composed of three kinds of layers: Convolution, pooling, and fully connected layers. In a CNN, the input image is convolved with a set of learnable filters or kernels that slide over the input image and extract spatial patterns or features. This process is known as convolution and the output of a convolutional layer is a set of feature maps that highlight different aspects of the input image. The feature maps are then passed through a nonlinear activation function, such as ReLU (Rectified Linear Unit), to introduce nonlinearity into the model. A CNN can be consists of multiple convolutional layers, followed by pooling layers that down sample the feature maps and reduce the spatial dimensions of the output. The pooled feature maps are then fed into one or more fully connected layers that learn to classify the image based on the extracted features. In Fig. 1 we show schematic representation of our architecture of CNN \cite{mat}. It consists of two CNN layers each layer composes of convolutional layer, activation layer (ReLU) and pooling layer. The two layers of CNN are followed by flattened layer and two fully connected (FC) layers. To measure the performance of our CNN, we use cross-entropy loss with soft-max loss function. The ADAM optimizer is used to speed up our neural network. 
\begin{figure}[htb]
\includegraphics[width=100mm,height=70mm]{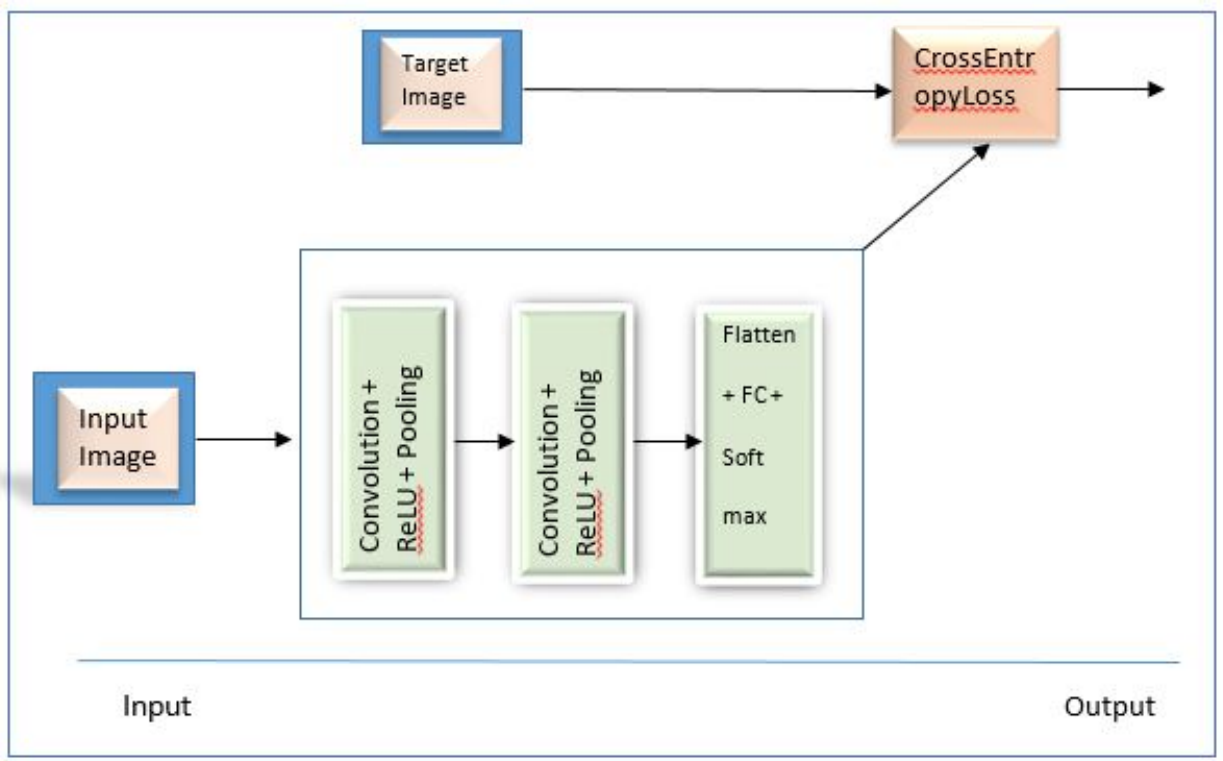}
\caption{Schematic representation of our CNN architecture.}
 \end{figure}

\subsection{Unsupervised Learning Algorithm}
The density-based spatial clustering of applications with noise (DBSCAN) is a density-based clustering algorithm commonly used in unsupervised learning. It was proposed by Martin Ester et. al. in 1996 \cite{est}. DBSCAN is widely used in various applications such as spatial data analysis, image segmentation, anomaly detection and customer segmentation. The advantages of DBSCAN include its ability to discover clusters of any arbitrary sizes and shapes, its ability to detect outliers as noise points and it does not require specifying the number of clusters in advance. However, it may struggle with datasets of varying densities and suffers from parameter sensitivity. The main idea behind DBSCAN is to group together data points that are close to each other in regions of high density while separating regions with lower density. Unlike traditional clustering algorithms like K-means, DBSCAN does not require specifying the number of clusters in advance and can discover clusters of various shapes. We can describe the method that DBSCAN algorithm works as follows \cite{est}:
\begin{itemize}
	\item Density-Based: DBSCAN defines clusters based on the density of data points. It considers a data point as a core point if there are at least a minimum number of data points (minPts) within a specified distance $\epsilon$ of that point are called $\epsilon$-neighborhood of a core point. Core points are the foundation of clusters.
	\item Directly Density-Reachable: A data point is directly density-reachable from a core point if that point belongs to the $\epsilon$-neighborhood of that core point and the $\epsilon$-neighborhood of the core point are larger than or equal minPts. 	
	\item Density-Reachable: Two data points are density-reachable if they can be reached by a chain of points (each within $\epsilon$ distance of the previous one), in which for every two consecutive points in the chain the later point is directly density-reachable to the preceding one.
	\item Density-connectivity: A data point is density connected to any other point, if there is a common point between those two points such that both of them are density-reachable to that common point.	
	\item Noise Points (Outliers): Data points which are not core points and cannot be reached by density connectivity are treated as noise points or outliers.
\end{itemize}
To find a cluster, DBSCAN algorithm starts randomly selecting a data point that has not been visited and retrieves all data points density-reachable from it. If the chosen point is a core point, this procedure yields a cluster, otherwise DBSCAN visits the next point of the database.

\section{Directed Percolation (DP) and Model Description}
Percolation occurs when the isolated clusters of networks change under the effect of system parameters into a fully connected structure whose size is of the order of the network size \cite{hi}. For example in the two dimensional bond percolation on the networks if we consider the connection between any two neighbors on the network is opened with probability $p$, hence for small values of $p$, there will be many small of isolated clusters in the network. However, when $p$ is increased gradually, these clusters start to merge and at a certain value of $p=pc$, called the percolation threshold, a giant connected component cluster spans the lattice. 
The change in the clusters size from isolated clusters to a spanning cluster marks a phase transition. There are two kinds of bond percolation, isotropic (undirected) bond percolation and anisotropic (directed) bond percolation. Anisotropic bond percolation which abbreviate as directed percolation (DP) involves the flow in a specific direction in space. The bonds (channels) work as valves in a way that the flow in the network can only percolate on a given direction.

In the dynamic systems if we consider the given direction as time, DP may be interpreted as $d+1$ dimensional system describes the spreading of some non-conserved process. Phase transition in this class of system occurs between the two distinct phases, the phase where the spreading on the network dies out (absorbing phase) and the phase where the spreading survives (active phase). DP phase transition is the most famous class of nonequilibrium phase transition. Many of systems have been found to be in this class. The good thing which help us to use machine learning to identify the phase transition of DP class is that: Even so simple numerical simulations of any system of DP kind can show the temporal evolution of such systems changes significantly at the phase transition. By means of that, from the typical space-time snapshot of this class, we can distinguish simply between the absorbing phase and active phase. In Fig. 2 we show typical space-time evolution of system of DP class in $(1+1)$ dimension. Top panel of Fig. 2 shows the case when we start simulations initially from fully occupied lattice, and bottom panel shows when we start simulations with single occupied site. It is clear from Fig. 2 that, the behavior of DP on both sides of $p_c$ is completely different. We can clearly from the images distinguish between the absorbing phase from active phase. Therefore, we will exploit this property to train CNN in order to become able to tell us to which phase the image belongs to when we feed it with new unseen image.
\begin{figure}[htb]
\includegraphics[width=150mm,height=100mm]{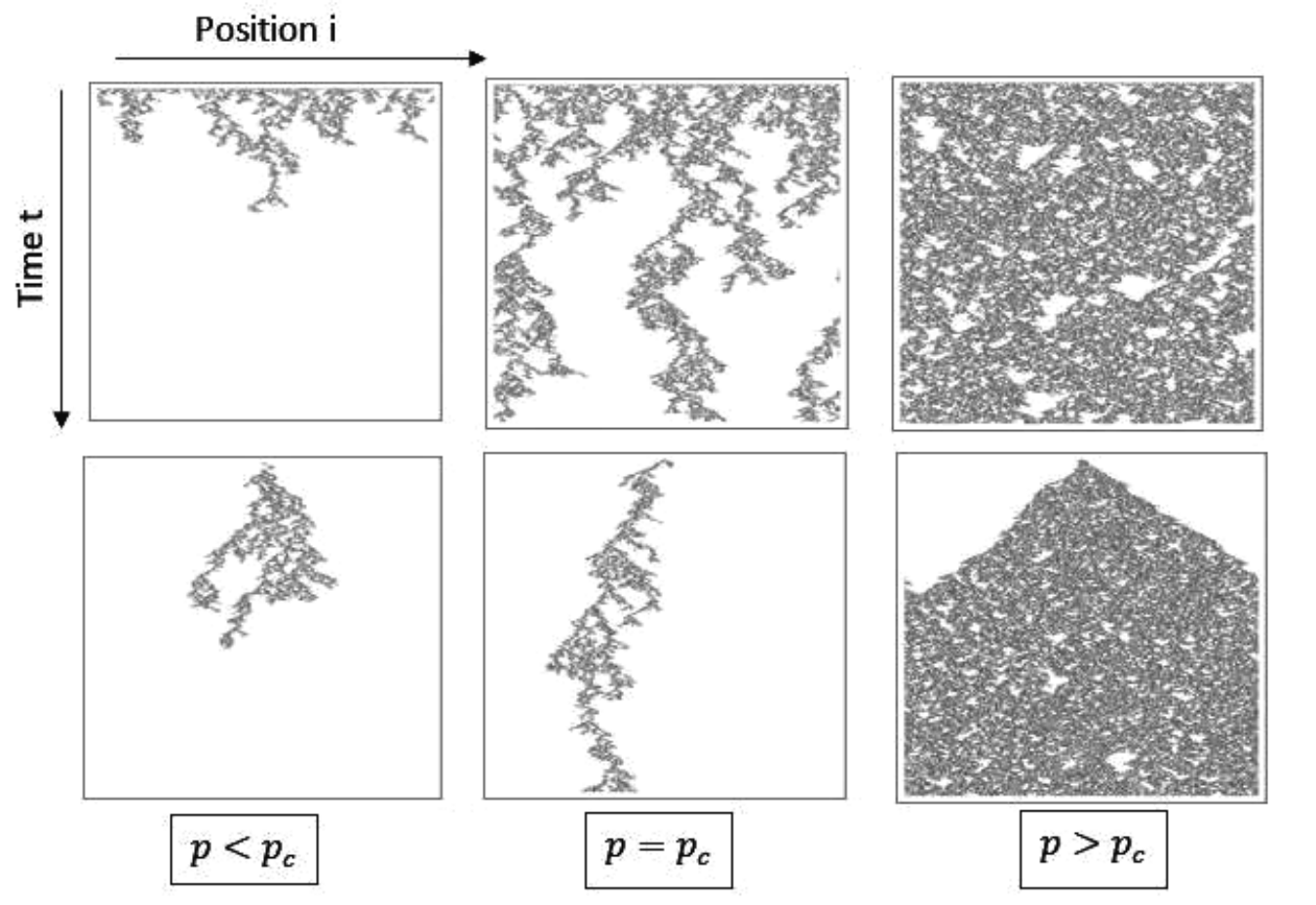}
\caption{Typical DP bond clusters in $(1+1)$ dimensions grown initially from: (top panel) fully occupied lattice, (bottom panel) a single active seed.}
 \end{figure}

Using CNN and DBSCAN algorithms, we aim to study the following two simple models in $(1+1)$ dimension. Both models have been proven to undergo a phase transition from the absorbing phase to active phase which belongs to DP class. In what follows, we give a brief description of those two models.
\subsection{Directed Bond Percolation} 
Directed bond percolation (bond DP) process on one dimension describes the time evolution of binary variable $s_i(t)$ of one dimensional lattice site, where $i$ denotes to spatial coordinate (horizontal axes) and t is a discrete time variable (vertical axes) \cite{hi}. The site is active (occupied) if $s_i=1$ and inactive (empty) when $s_i=0$. Time evolution of this model occurs according to the following rules:
\begin{eqnarray}
s_i(t+1)=\left\{\begin{array}{lllll} 
  1 & \mbox{if}& s_{i-1}=1& \mbox{and}&z^-<p\\ 
  1 & \mbox{if}& s_{i+1}=1& \mbox{and}&z^+<p\\                             
  0 &  \mbox{otherwise}                   
        \end{array}
        \right.
\end{eqnarray}
Where $p$ is the probability for the bond between any two connected sites to be open. $z^\pm\in[0,1]$ is random number selected from uniform distribution. This model has been found to show a DP phase transition with critical point at $p_c=0.6447$ \cite{je}.

\subsection{The Domany-Kinzel Cellular Automaton}
Second model which we intend to study using machine learning is the Domany-Kinzel cellular automaton (DK) on $(1+1)$ dimension \cite{hi}. The sites $i$ of DK model at any time take the binary value $s_i(t)=\left\{0,1\right\}$. The time evolution of the DK model occurs according to the following rules:
\begin{eqnarray}
s_i(t+1)=\left\{\begin{array}{lllll} 
  1 & \mbox{if}& s_{i-1}\neq s_{i+1}& \mbox{and}&z_i(t)<p_1\\ 
  1 & \mbox{if}& s_{i-1}=s_{i+1}=1& \mbox{and}&z_i(t)<p_2\\                             
  0 &  \mbox{otherwise}                   
        \end{array}
        \right.
\end{eqnarray}
In contrast to bond DP, the DK model depends on two percolation probabilities $p_1$ and $p_2$. $z_i(t)\in[0,1]$ is random number selected from uniform distribution. This model has been found to undergo a phase transition of DP class at the critical point $p_1=0.64770$ and $p_2=87376$ \cite{hi}.

\section{Learning Percolation by CNN Method}  
Our goal on this work to train the CNN on large numbers of images which are similar to the images in Fig. 2. After the training the CNN should be able to decide to which phase the images belong when feeds with a new unseen image. We have two choice in hand to learn machine. The first choice to use directly the raw data of Monte Carlo simulations to train machine, whereas the second choice is to start converting the raw data of Monte Carlo simulations to images, after that we use those images to train machine. Whereas the second choice needs more efforts we adopt it here for the following reasons:  
\begin{itemize}
\item We need to train the machine only on the images generated from a one model, e. g. bond DP, and use that trained machine to study any new model of DP class even if we do not have a prior knowledge of the value of critical point for that model. 
	\item We do not need to retrain the machine at each time we change the lattice size or simulation time. 
\end{itemize}

For purpose of using CNN algorithm to determine the critical points of any system of DP universality class, we start training it using the images which have been generated from Monte Carlo simulations of bond DP model. Each image used in training is due to the Monte Carlo simulation of lattice of size $L=1000$ sites, and for time is $T=1000$ time steps. Hence, the dimensions of each image is $L \times T$. We have generated (training images) $150$ images in subcritical region ($p<P_c$) and $150$ images in supercritical region ($p>p_c$). we give the label (0) for the images in subcritical region and (1) for the images in supercritical region. We have also generated $60$ images for validation (testing images). Close to the critical point we consider the image to be of label 1 if that image survives for time which is greater than the correlation time at that point such the image in Fig. 3 (a), otherwise the image label is 0 such as the image in Fig. 3 (b). During the training, machine encodes each image to $100 \times 100$ pixels in size with color space is gray scale. We find that, this value of pixels is enough to get a good performance for machine, however smaller values of pixels do not work properly. Fig. 4 shows an example of original image and its encoded images. After we train the CNN using the images described above we find that: Even with this small sample of images CNN can predict the label of any image with accuracy of $99\%$ after $8$ epoches of training. This value of accuracy does not mean the CNN can determine the critital points with that accuracy, however it means the CNN can predict the label of the images with that value of accuracy.  Ability of CNN in determantion of critical points will depend on our selection of the label for the images near the critical point which we use it in training. As we have shown in Fig. 3, there is some difficulty in determantion the label of the images as we approach the critical point.  
\begin{figure}[htb]
\includegraphics[width=80mm,height=40mm]{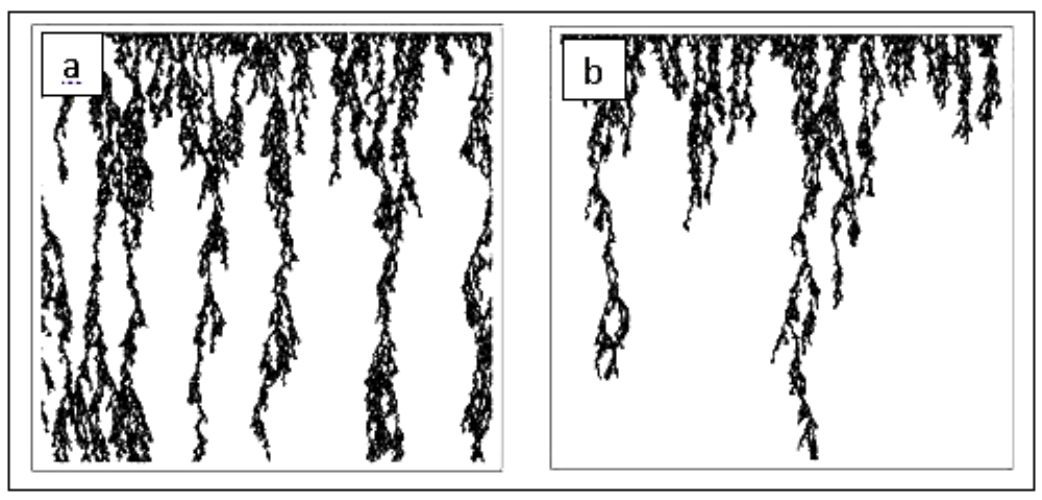}
\caption{Generated images close to the critical point.}
 \end{figure}

\begin{figure}[htb]
\includegraphics[width=120mm,height=40mm]{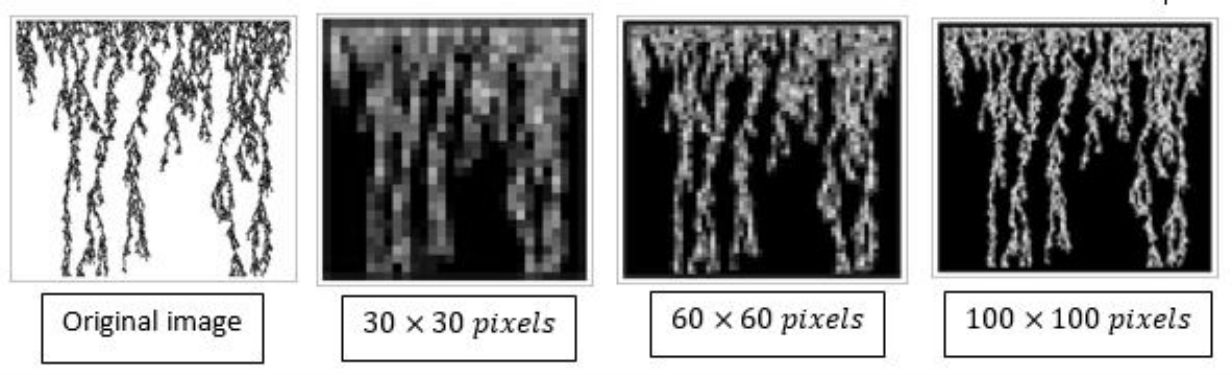}
\caption{Original image and its encoded images.}
 \end{figure} 
     
We use the trained CNN in order to determine the critical points for the both models, bond DP and DK, at various values of lattice size ($L=20, 40, 60, 80, 100$). The dimension of each image which we generate is always $L \times T$. To take the effect of correlation time in account we set $T$ to be proportional to $L^Z$, where $z=1.580(1)$ is the dynamical exponent \cite{hi}. During the testing, machine encodes each image to $100 \times 100$ pixels in size. Fig. 5 (a) shows the average probability of being the label of image is 0 ($P_0$) as function of $p$ for lattice of size $L=20$ for bond DP model. We consider the point $p$ where the CNN predicts 0 and 1 with equal probability to be the critical point $p_c$ for the model \cite{xc,she}. Estimated value for the critical point in this case as we can see from Fig. 5 (a) is $p_c=0.619(2)$. For the case of DK model, we fix the value of the first probability $p_1$ to be equal to the critical point of this model $p_{1c}=0.6447$ \cite{hi}, and allow CNN to determine the second probability $p_{2c}$. Fig. 5 (b) shows $P_0$ for DK model as function of $p_2$ for lattice of size $L=20$.  The critical point estimated in this case is $p_{2c}=0.792(7)$ see Fig. 5 (b). For both models, we averaged over $100$ images far from critical point and $500$ images beside the critical point. 
\begin{figure}[htb]
\includegraphics[width=70mm,height=60mm]{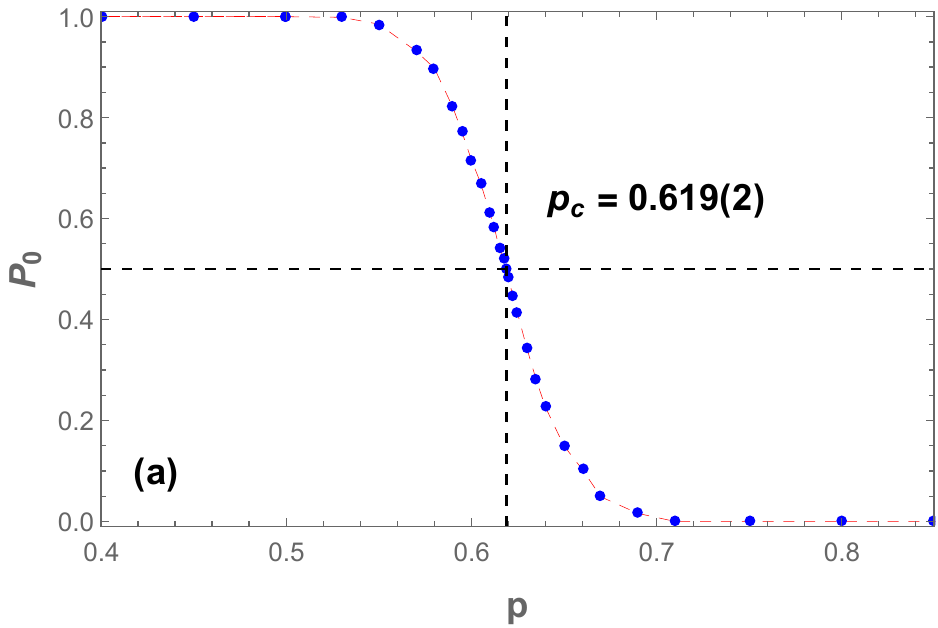}
\includegraphics[width=70mm,height=60mm]{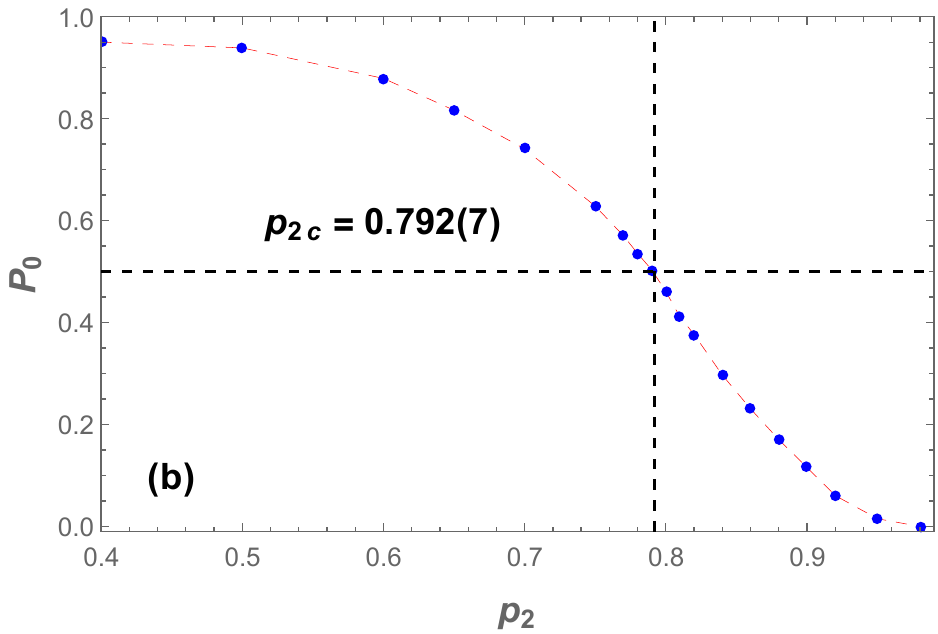}
\caption{CNN results of one-dimensional with image configuration as input to CNN at $L=20$. (a) Bond DP, (b) DK model.}
 \end{figure}  
  
In order to determine the critical points for both models when the system size goes to infinity, we extract the critical points using CNN at various values of lattices sizes for both models. For the case of bond DP model, Fig. 6 (a) shows the critical points $p_c$ predicted by CNN as function of $1/L$ for the values of $L=20, 40, 60, 80$ and $100$. Extrapolation of the results to an infinite system size suggests the critical point for bond DP in $(1+1)$ dimension to be $p_c=0.6454(5)$, which agrees very well with the standard value of $0.6447$ \cite{je} for this model. The estimated value for critical point we find here is exactly the value of critical point that have been found for this model using DANN \cite{she}. For DK model we fix again the value of $p_{1c}$ to be $0.6447$ and use CNN to predict the values of $p_{2c}$. In Fig. 6 (b) we plot the values of the critical points $p_{2c}$ predicted by CNN as function of $1/L$ for the lattices of sizes $L=20, 40, 60, 80$ and $100$. Estimated critical point in this case is $p_{2c}=0.8768(4)$ in the limit of an infinite system size. This value again coincide very will with the standard value of $0.87376$ \cite{hi} for this model.
\begin{figure}[htb]
\includegraphics[width=70mm,height=60mm]{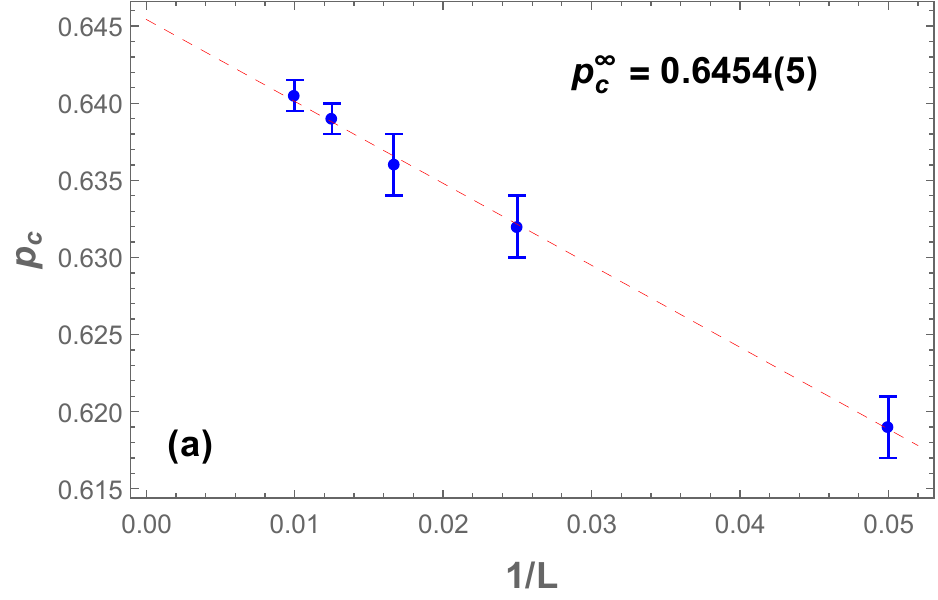}
\includegraphics[width=70mm,height=60mm]{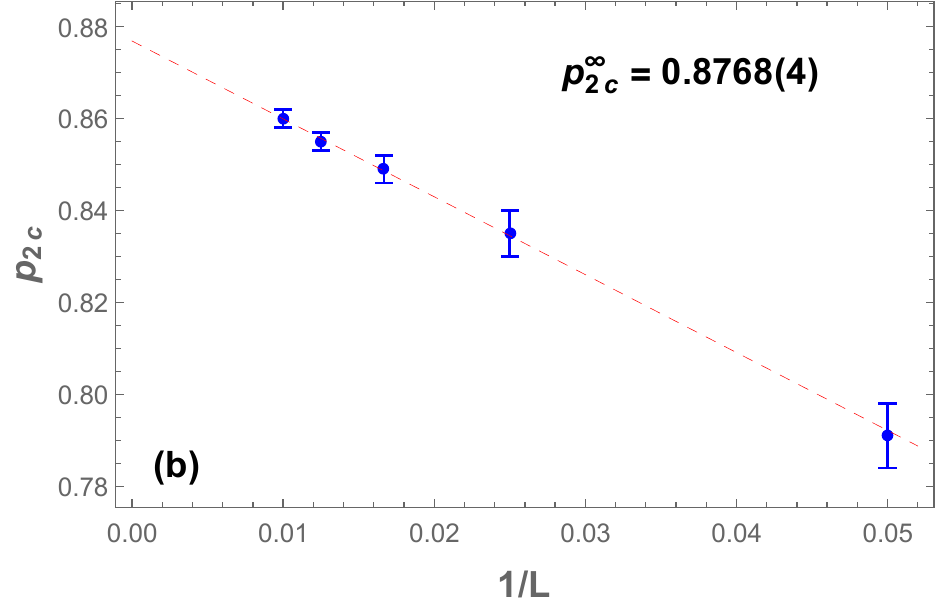}
\caption{Critical points as function of $1/L$ for the values of $L=20, 40, 60, 80$ and $100$ using CNN for $(1+1)$ dimension of: (a) Bond DP, (b) DK model.}
 \end{figure}
 
\section{Learning Percolation by DBSCAN method} 
To train machine using DBSCAN algorithm for both models bond DP and DK, we generate a samples of $n$ uncorrelated configurations for each value of probabilities which are from $0.1$ to $1$ with an interval of $0.1$ using the Monte Carlo simulation. Each configuration is taken after updating the model for time $t>L^Z$, where $z=1.580(1)$ is the dynamical exponent \cite{hi}. We collect the samples into an $M \times L$ matrix,
\begin{eqnarray}
X=\left(\begin{array}{lllll}
0&1&...&1&0\\
 & & . & & \\
 & & . & & \\
1&1&...&0&1\\
 \end{array}\right)_{M \times L}
\end{eqnarray}
where $M=10n$ is the total number of configurations, and $L$ is the number of lattice sites. Each element $X_{ij}$ in the matrix $X$ describes the state of site $j$ on the configuration $i$. Such a matrix of raw data is the only data we feed to the DBSCAN algorithm. The learning with unsupervised does not need a prior knowledge of the values of the critical points unlike with learning using supervised machine. Here we use the machine to extract prominent features of the data and then use this information to classify the samples into distinct phases.

Initially and before we go to use DBSCAN to study the phase transition, we would like to know how our configurations distributed in two dimensional space. For that we will reduce the dimensional space of our configurations from the dimension $L$ to two dimensional space. There are many of unsupervised learning algorithms perform that task. Latent semantic analysis (LSA), t-distributed stochastic neighbor embedding (t-SNE), a trainable autoencoder, locally linear embedding (LLE), principle components analysis (PCA) and isometric mapping are examples of unsupervised learning algorithms which project the high-dimensional data into a lower-dimensional approximating manifold. Here our purpose is mainly to show the configurations distribution, so LSA algorithm is a suitable tool to execute that. Fig. 7 shows the clusters distributions for bond DP model with lattice of size $L=20$ for $M=4000$ configurations after we update the model for $10$ time steps (left) and $40$ time steps (right). In that figure we use LSA to reduce the dimension of our configurations from $20$ to $2$ and use DBSCAN algorithm for classification. It is clear from the figure that, for small values of updating time DBSCAN classifies the system to three classes (left of Fig. 7), that means the informations used to train DBSCAN are not sufficient. However as we update the system for longer time DBSCAN becomes able to predict the correct labeling (right of Fig. 7).

\begin{figure}[htb]
\includegraphics[width=40mm,height=40mm]{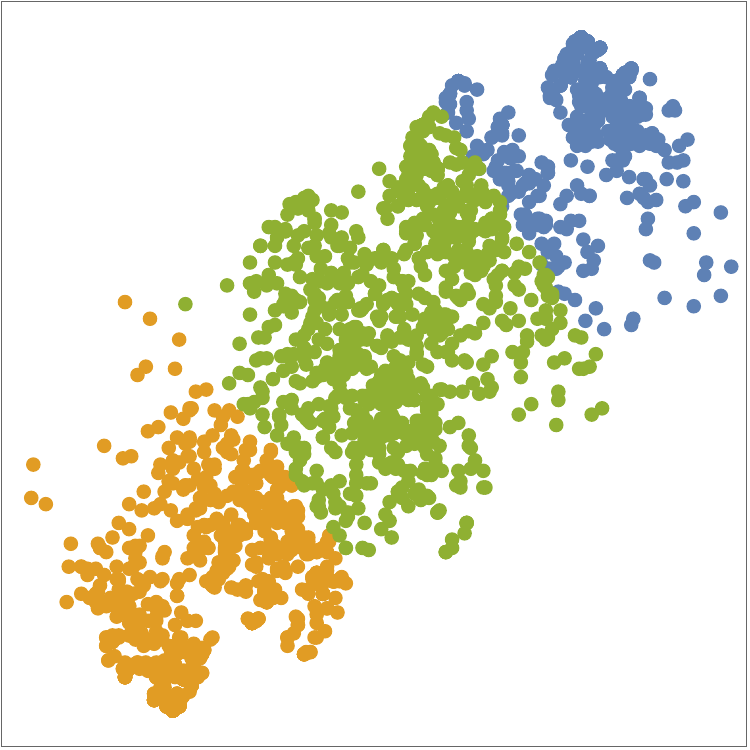}
\includegraphics[width=40mm,height=40mm]{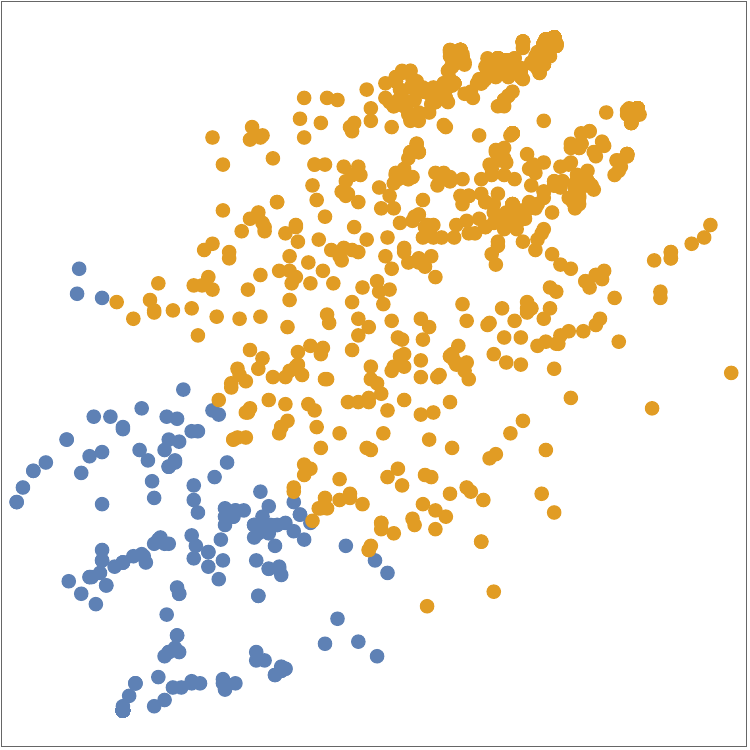}
\caption{Projection of the configurations onto the two dimensional plane for bond DP model on one dimension lattice of size $L=20$ after $10$ steps of iterations (left) and $40$ iterations (right) for $M=2000$ configurations.}
 \end{figure}
 
To determine the critical points accurately, we use Monte Carlo simulations to prepare a matrix $X$ Eq. 3 for both models, bond DP model and DK model. Our matrix $X$ consists of $M=1000$ configurations have been prepared as we describe in the beginning of this section. Training of DBSCAN algorithm with the matrix $X$ reveals that, the DBSCAN classifies the configurations $M$ automatically into two classes. We use the trained DBSCAN to study the phase transition for both models bond DP and DK. The results of using DBSCAN to determine the critical points for lattice size $L=20$ is shown in Fig. 8. In that figure, we plot the average probability $P_0$ of being the label of the configuration is 0 as function of $p$ for bond DP model (Fig. 8 (a)) and $p_2$ for DK model (Fig. 8 (b)). For DK model we fix the value of $p_1=0.6447$ as in CNN case. 

\begin{figure}[htb]
\includegraphics[width=70mm,height=60mm]{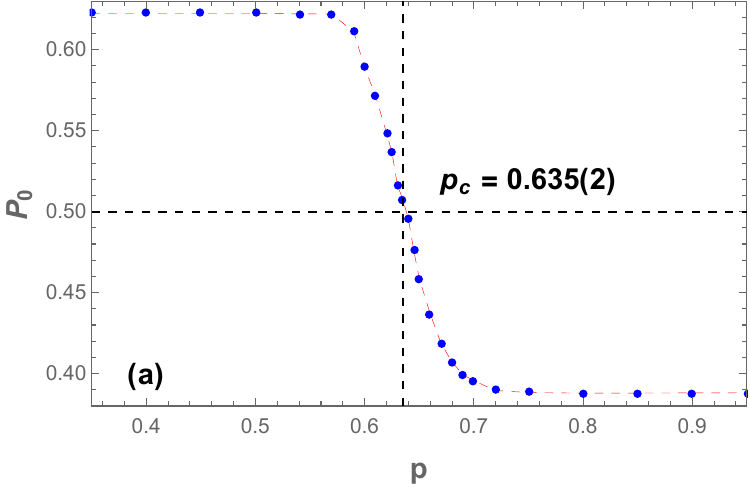}
\includegraphics[width=70mm,height=60mm]{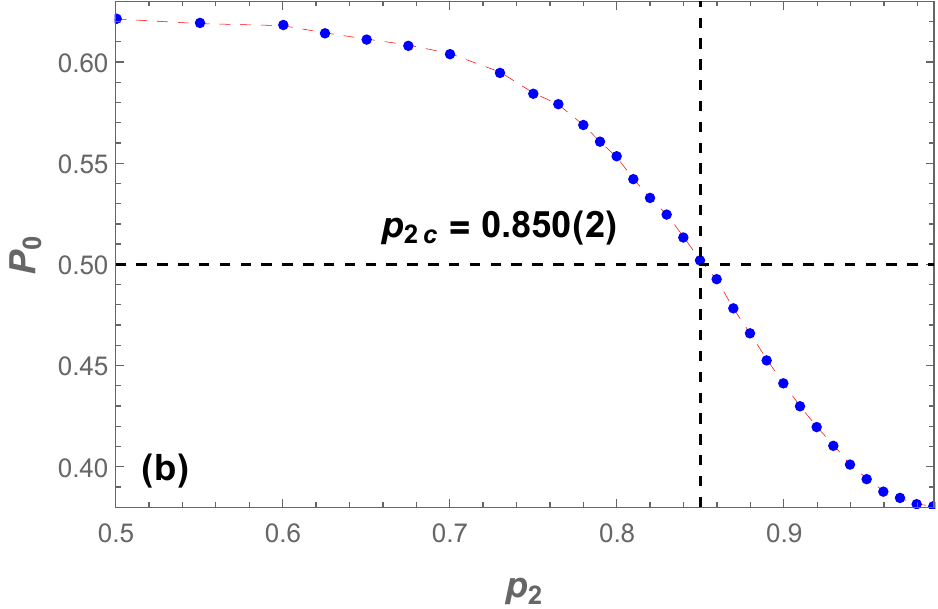}
\caption{DBSCAN results of one-dimensional with raw data configurations as input to DBSCAN at $L=20$. (a) Bond DP, (b) DK model.}
 \end{figure}

To get the critical points for both models we use DBSCAN to extract further critical points for lattices of sizes $L=20, 40, 60, 80$ and $120$. Fig. 9 shows the critical points which we have obtained using DBSCAN at various values of lattice size for both models. Extrapolation of the critical point to infinite lattice size for bond DP model suggests to be $p_c=0.6455(2)$ and for DK model to be $p_{2c}=0.8741(2)$. These values of critical points we find here coincide exactly with those points we found using CNN. Table 1 summarizes the critical points for bond DP model and DK model which we get them using CNN algorithm and DBSCAN algorithm comparing to the results which have been found for the critical points for the same models in previous studies. We can note that, the value of critical point that we find it here for bond DP using CNN and DBSCAN is the same value that has been found using DANN \cite{she}. Finally the previous results with the results have been obtained in preceding studies \cite{zha,she1,she} confirm the ability of machine learning algorithms in helping to determine accurately the critical points for models which have a phase transition of DP universality class.  
  
\begin{figure}[htb]
\includegraphics[width=70mm,height=60mm]{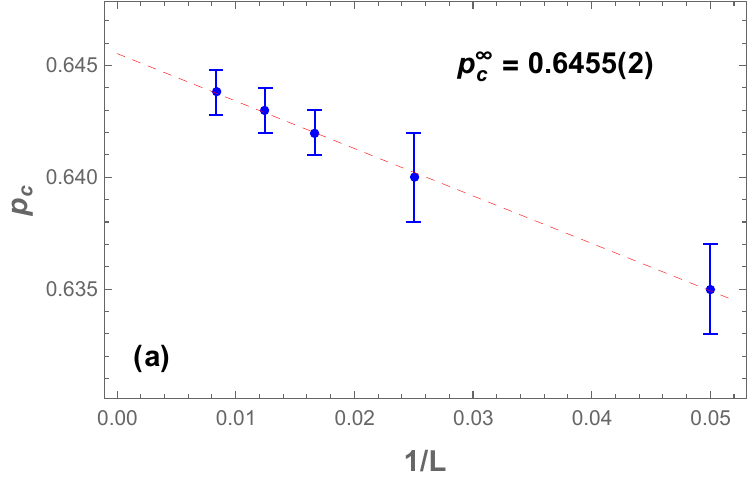}
\includegraphics[width=70mm,height=60mm]{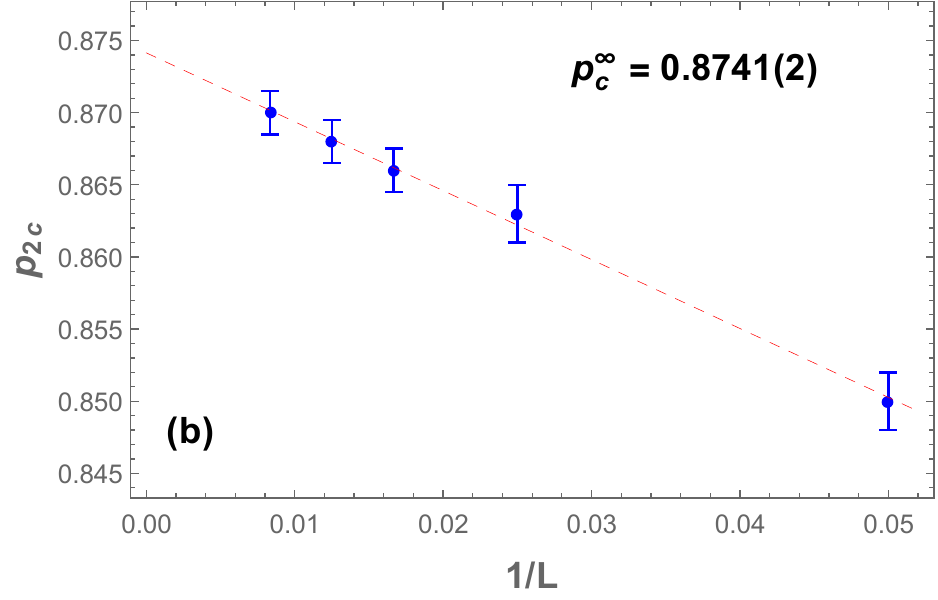}
\caption{Critical points as function of $1/L$ for the values of $L=20, 40, 60, 80$ and $120$ using DBSCAN for $(1+1)$ dimension of: (a) Bond DP, (b) DK model.}
 \end{figure}

\begin{tabular}{|c|c|c|c|c|}
\multicolumn{5}{l}{Table 1: Critical points for bond DP and DK models.}\\
\hline
 &CNN (this work)&DBSCAN (this work)& DANN \cite{she}& Standard \cite{hi}\\\hline
Bond DP&0.6454(5)&0.6455(2)&0.6453(5)&0.64770\\\hline
DK model&0.8768(4)&0.8741(2)&--&0.87376\\\hline
\end{tabular}\\

\section{Conclusion} 
Using machine learning with supervised learning with CNN algorithm and unsupervised learning with DBSCAN algorithm, we have determined the critical points for bond DP model and DK model. The critical points which we find using CNN and DBSCAN for both models are the same values of the critical points those have been obtained using Monte Carlo simulations methods. We assert that, unlike the Monte Carlo methods which need to be applied on a large lattice size to get accurately the critical points of models under study, machine learning can work even with a very small lattice size and predicts the critical points accurately. Therefore, beyond a mean field theory and Monte Carlo simulations, machine learning algorithms can be considered as a new tool in hand to deal with phase transitions in equilibrium and nonequilibrium systems.



\begin{thebibliography}{35}

\bibitem{mi}R. S. Michalski, J. G. Carbonell and T. M. Mitchell, Machine Learning An Artificial Intelligence Approach (Springer-Verlag Berlin Heidelberg GmbH 1984).
 
\bibitem{goo}I. Goodfellow, Y. Bengio and A. Courville, Deep Learning (MIT Press Cambridge, MA, 2016), Vol. 1.

\bibitem{eng} A. Engel and C. Van den Broeck, Statistical Mechanics of Learning (Cambridge University Press, Cambridge, UK, 2001).

\bibitem{gor}M. I. Jordan and T. M. Mitchell, Science {\bf 349}, 255 (2015).

\bibitem{tag}L. Tagliaferri, M. Morales, E. Birbeck and A. Wan, Python Machine Learning Projects (DigitalOcean, New York City, New York, USA).

\bibitem{yam}R. Yamashita1, M. Nishio1, R. K. Gian Do and K. Togashi, Insights into Imaging {\bf 9}, 611 (2018).

\bibitem{car}G. Carleo, I. Cirac, K. Cranmer, L. Daudet, M. Schuld, N. Tishby, L. Vogt-Maranto, and L. Zdeborov$\acute{a}$, Rev. Mod. Phys. {\bf 91}, 045002 (2019).

\bibitem{zde}L. Zdeborov$\acute{a}$, Nat. Phys. {\bf 13}, 420 (2017).

\bibitem{wan}L. Wang, Phys. Rev. B {\bf 94}, 195105 (2016).

\bibitem{hua}S. Huang, W. Klein and H. Gould, Phys. Rev. E {\bf 103}, 033305 (2021).

\bibitem{zha}W. Zhang, J. Liu and T.-C. Wei, Phys. Rev. E {\bf 99}, 032142 (2019).

\bibitem{li}C. Li, D. Tan and F. Jiang, Annals of Physics {\bf 391}, 312 (2018).

\bibitem{tan}D. Tan, C. Li, W. Zhu and F. Jiang, New Journal of Physics {\bf 22}, 063016 (2020).

\bibitem{xc}X. Chena, F. Liua, S. Chena, J. Shend, W. Deng, G. Pappb, W. Lia and C. Yang, Physica A {\bf 617}, 128666 (2023).

\bibitem{chn}K. Chng, N. Vazquez and E. Khatami, Phys. Rev. E {\bf 97}, 013306 (2018).

\bibitem{car1}J. Carrasquilla and R. G. Melko, Nat. Phys. {\bf 13}, 431 (2017).

\bibitem{sc}S. S. Schoenholz, E. D. Cubuk, D. M. Sussman, E. Kaxiras and A. J. Liu, Nat. Phys. {\bf 12}, 469 (2016).

\bibitem{ven}J. Venderley, V. Khemani and Eun-Ah Kim, Phys. Rev. Lett. {\bf 120}, 257204 (2018). 

\bibitem{don}X. -Y. Dong, F. Pollmann and X. -F. Zhang, Phys. Rev. B {\bf 99}, 121104(R) (2019).

\bibitem{hue}P. Huembeli, A. Dauphin and P. Wittek, Phys. Rev. B {\bf 97}, 134109 (2018).

\bibitem{tom}Y. Tomita, K. Shiina, Y. Okabe and H. K. Lee, Phys. Rev. E {\bf 102}, 021302 (2020).

\bibitem{she1}J. Shen, W. Li, S. Deng and T. Zhang, Phys. Rev. E {\bf 103}, 052140 (2021).

\bibitem{she}J. Shen, F. Liu , S. Chen , Dian Xu, X. Chen, S. Deng, Wei Li, G. Papp and C. Yang, Phys. Rev. E {\bf 105}, 064139 (2022). 

\bibitem{van}E. P. Van Nieuwenburg, Y. -H. Liu and S. D. Huber, Nat. Phys. {\bf 13}, 435 (2017).

\bibitem{hu}W. Hu, R. R. P. Singh and R. T. Scalettar, Phys. Rev. E {\bf 95}, 062122 (2017).

\bibitem{we}S. J. Wetzel, Phys. Rev. E {\bf 96}, 022140 (2017).

\bibitem{wa}J. Wang, W. Zhang, T. Hua and T. -C. Wei, Phys. Rev. Research {\bf 3}, 013074 (2021).

\bibitem{wet}S. J. Wetzel, Phys. Rev. E {\bf 96}, 022140 (2017).

\bibitem{lec}Y. Lecun, L. Bottou, Y. Bengio and P. Haffner, Proc IEEE {\bf 86}, 2278 (1998).

\bibitem{kr}A. Krizhevsky, I. Sutskever and G. E. Hinton, Advances in Neural Information Processing Systems,  pp. 1097 (2012).

\bibitem{goh}A. Ghosh, Abu Sufian, F. Sultana, A. Chakrabarti and Debashis De, \href{https://www.researchgate.net/publication/337401161}, (2020).

\bibitem{mat}Neural Networks in the Wolfram Language, help of Mathematica 12.0 program.

\bibitem{est}M. Ester, H. P. Krigel, J. Sander and X. Xu, Proc. of 2nd International Conference on Knowledge Discovery and Data Mining (KDD-96), WA, pp. 226-231 (1996).

\bibitem{hi}H. Hinrichsen, Adv. Phys. {\bf 49}, 815 (2000).

\bibitem{je}I. Jensen, J. Phys. A: Math. Gen. {\bf 37}, 6899 (2004). 
 
\end{thebibliography}
\end{document}